\documentclass[usenatbib]{mn2e}
\usepackage{graphics}

\begin{document}

\title[Marked statistics]{Breaking Halo Occupation Degeneracies
with Marked Statistics}

\author[White \& Padmanabhan]{
Martin White$^{1,2}$, Nikhil Padmanabhan$^{2}$\\
$^{1}$ Departments of Physics and Astronomy, University of California,
Berkeley, CA 94720, USA\\
$^{2}$ Lawrence Berkeley National Laboratory, 1 Cyclotron Road, Berkeley,
CA 94720, USA}

\date{\today} 
\pagerange{\pageref{firstpage}--\pageref{lastpage}}

\maketitle
\label{firstpage}

\begin{abstract}
We show that a suitably defined marked correlation function can be used to
break degeneracies in halo-occupation distribution modeling.  The statistic
can be computed on both 3D and 2D data sets, and should be applicable to all
upcoming galaxy surveys.  A proof of principle, using mock catalogs created
from N-body simulations, is given.
\end{abstract}

\begin{keywords}
cosmology: large-scale structure
\end{keywords}

\section{Introduction}

In recent years our ability to describe galaxy clustering has advanced
dramatically.
The halo model (\citealt{Sel00,PeaSmi00}, see e.g.~\citealt{CooShe} for
a review) has provided us with a physically informative and flexible means
of describing galaxy bias - the relation between galaxies and the underlying
dark matter halos.
The key insight is that an accurate prediction of galaxy clustering
requires knowledge of how galaxies are apportioned between and
distributed within halos - the halo occupation distribution or
HOD. Combined with the theoretically predicted spatial distribution of
halos from e.g.~N-body simulations, a specified HOD makes strong
predictions about a wide array of galaxy clustering statistics.  The
formalism is now widely used in the interpretation of galaxy
clustering and to infer cosmological parameters from large-scale
galaxy surveys.

Much of the recent work on fitting HODs has used the two-point galaxy 
correlation function as the observation of choice. While the galaxy correlation
function provides very strong constraints on the HOD, there exist degeneracies
between the inferred HOD and the underlying cosmology. 
Much of this degeneracy arises because
a change in the cosmology, and hence the halo population, can be compensated
to a large extent by a change in the galaxy halo occupation.  This modified
halo occupation apportions galaxies differently amongst halos of different mass
than the fiducial model.  Not surprisingly, combining the galaxy correlation 
function with a second observable with different sensitivities to the HOD 
can lift such degeneracies \citep{ZheWei07} tightening the constraints and
allowing one to simultaneously constrain the cosmological world model and HOD
\cite[see e.g.][]{Aba05}.
A number of such observables have been considered in the literature.
Galaxy-galaxy lensing has the potential to directly measure the mass of 
the halos hosting a galaxy population. 
Cross-correlating with another galaxy sample selected to live in
high (or low) mass halos can help to break degeneracies, as can redshift
space distortions (which are very sensitive to the satellite fraction) or
peculiar velocities.
Another observation is the abundance of rich clusters of galaxies, which can
constrain the number density of massive halos.
While all of these approaches are certainly valid, and will continue to be
used in the future, a disadvantage is that they generally require additional
observations or measurements, with the associated modeling and additional
systematics that must be calibrated.
A natural question, therefore, is whether one can break these degeneracies
using only the data going into the clustering statistics themselves?

Higher order clustering measurements provide one such approach
\citep[e.g.][]{Kul07}.
A particularly convenient choice for our purposes is a marked correlation
function, where the mark is determined from the galaxy spatial distribution.
While this represents new information, it is information which is readily
available.
We demonstrate here that appropriately chosen marked two-point correlation
functions
\citep{BeiKer00,BeiKerMec02,Got02,SheTor04,SheConSki05,Har06,Wec06}
can lift degeneracies in the HOD or between the HOD and the cosmology.
Such marked correlation functions are straightforward to compute with the
same set of observations, and require no additional data nor understanding
of the survey or algorithms beyond those required for basic clustering
statistics.
This paper serves as a proof-of-principle, by demonstrating that the
degeneracy between HOD and the amplitude of the primordial fluctuation
spectrum is broken with a simple density mark in simulations.

\section{A Worked Example}

\subsection{Degeneracies}

\begin{figure}
\begin{center}
\resizebox{2.75in}{!}{\includegraphics{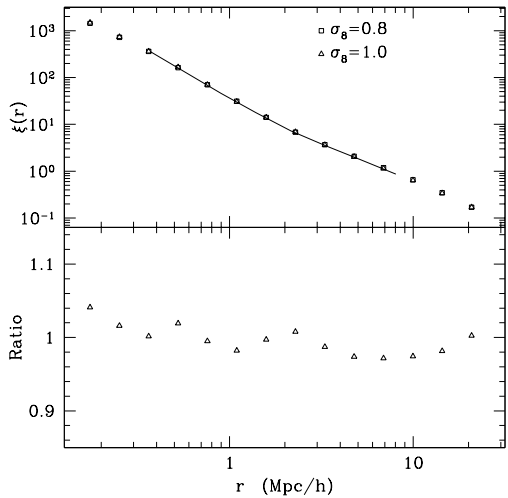}}
\end{center}
\caption{(Top) The best-fit correlation function from our two cosmologies with
$\sigma_8=0.8$ (open squares) and $\sigma_8=1.0$ (open triangles) along with
the input `data' (solid line).  The line extends only over the range of the
fit, and errors (between 5-10\%) are suppressed for clarity.
(Bottom) The ratio of the $\sigma_8=1$ correlation function to that of
$\sigma_8=0.8$.  Note that $\xi(r)$ for the two cosmologies is almost
identical, and well within our assumed errors.}
\label{fig:xir}
\end{figure}

\begin{figure}
\begin{center}
\resizebox{2.75in}{!}{\includegraphics{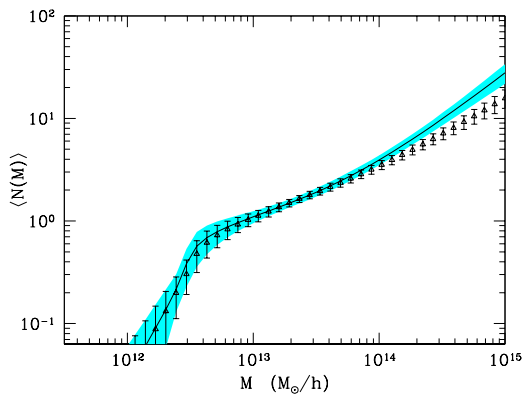}}
\end{center}
\caption{The HODs for the two cosmologies:
$\sigma_8=0.8$ (shaded region) and $\sigma_8=1.0$ (open triangles).
The error bars and width of the shaded region show the standard deviation from
elements of a Markov-chain Monte-Carlo.  The HODs differ by $\sim 2\sigma$ at
high mass.}
\label{fig:hod}
\end{figure}

To illustrate our point we consider a mock galaxy sample, with
characteristics similar to an $L_\star$ sample, at $z\sim 0.1$,
and try to analyze this in two cosmologies that only differ in the
normalization of the primordial power spectrum: $\sigma_8=0.8$ and $1.0$.
In both cosmologies a good fit to the 2-point function can be found
(Fig.~\ref{fig:xir}), but the HOD differs (Fig.~\ref{fig:hod})
because the halo mass function is different in the two cosmologies.

The fiducial galaxy sample was generated, and the fits were done, by populating
N-body simulations with galaxies using an HOD prescription.
We use a halo model which distinguishes between central
and satellite galaxies with a mean occupancy of halos:
$N(M)\equiv\left\langle N_{\rm gal}(M_{\rm halo})\right\rangle$.
Each halo either hosts a central galaxy or does not, while the number of
satellites is Poisson distributed about a mean $N_{\rm sat}$.
We parameterize $N(M)=N_{\rm cen}+N_{\rm sat}$ with 5 parameters
\citep[e.g.][]{Zheng05}
\begin{equation}
  N_{\rm cen}(M) = \frac{1}{2}
  \ {\rm erfc}\left[\frac{\ln(M_{\rm cut}/M)}{\sqrt{2}\sigma}\right]
\label{eqn:ncen}
\end{equation}
and
\begin{equation}
  N_{\rm sat}(M) = \left(\frac{M-\kappa M_{\rm cut}}{M_1}\right)^\alpha
\label{eqn:nsat}
\end{equation}
for $M>M_{\rm cut}$ and zero otherwise.  Different functional forms have
been proposed in the literature, but the current form is flexible enough
for our purposes here.

The fiducial galaxy sample is generated from the $\sigma_8=0.8$ simulation.
It has a number density of $1.5\times 10^{-3}\,h^3\,{\rm Mpc}^{-3}$ and a
correlation length of about $7\,h^{-1}$Mpc.
All errors are computed by Monte-Carlo methods, dividing the simulation
into disjoint regions.  For definiteness we consider a survey of volume
$(250\,h^{-1}{\rm Mpc})^3\simeq 1.6\times 10^7\,h^{-3}{\rm Mpc}^3$,
similar to the corresponding
Sloan Digital Sky Survey sample, and scale the covariance matrices to that
volume.  This yields diagonal errors on $\xi(r)$ of around $5-10\%$ and
bin-to-bin correlations\footnote{We use the full covariance matrix when
quoting significance levels.} of 15-80\%.
When fitting HOD models to these data the best fits are ``good'' fits, and
the parameter values are well within the range of HOD parameters seen for
similar galaxy samples, and so both cosmologies are acceptable {\it a priori}.
  
It is clear (Fig.~\ref{fig:xir}) that the two-point correlation function by
itself cannot distinguish between the two models - $\Delta\chi^2<1$ for 8
data points.
The next sections demonstrate that a simple density mark, measurable from the
spatial distribution of galaxies strongly discriminates between these models.

\subsection{Marked correlation functions}

The marked correlation function generalizes the standard correlation 
function by weighting galaxies by a numerical ``mark''. If the
mark of the $i^{th}$ object is $m_{i}$, then the marked correlation 
function is defined as 
\citep[e.g.][Eq.~3]{SheConSki05}
\begin{equation}
  M(r) = \frac{1}{ n(r)\bar{m}^2} \sum_{ij} m_i m_j ,  \;
\label{eqn:marked}
\end{equation}
where the sum is over all pairs of objects $(i,j)$ with separation $r_{ij}=r$,
$n(r)$ is the number of pairs, and the mean mark, $\bar{m}$, is calculated
over all objects in the sample.  Note that, unlike $\xi$, $w_p$ or $w$, no
random catalog is needed in the computation of $M(r)$.
It is convenient to divide out the clustering of the average sample since
$M(r)\ne 1$ then implies a difference in clustering by objects with different
marks.
The above expression can be applied in 2D or 3D, with angular or linear
bins.

The choice of mark depends on the application.  In the example above,
we would like a mark that encodes information about which halos host
which galaxies.
One set of such marks would be functions of the local density, which 
can be computed in a number of ways,
e.g.~the distance to the $n^{\rm th}$ nearest neighbor, the number of neighbors
within a fixed metric aperture, spline kernel interpolation
\citep[e.g.][]{Deh01} or kernel deprojection \citep[e.g.][]{Eis03}.
Massive halos tend to host more galaxies with higher density than lower
mass halos, so if the HOD is changed we expect the density-marked correlation
function to differ.
The relation of local density to the number density of groups or clusters
(which are known to break degeneracies in model fitting,
 e.g.~\citealt{ZheWei07}) can be complex, but is easily calculable from a
mock catalog.

What function of $\rho$ should we choose as our mark?  The choice is arbitrary,
but $\rho^n/(\rho_\star^n+\rho^n)$ has the nice property that it tends to zero
for $\rho\ll\rho_\star$ and unity for $\rho\gg\rho_\star$, the rapidity of the
transition being controlled by $n$.  This means the dynamic range in the mark
is limited, which leads to more stable results.  If we are concerned that our
density estimator may be noisy, which is often true in practice, we should
choose a low value of $n$.  Hereafter we choose $n=1$.

\subsection{Different HODs, different marks}

\begin{figure}
\begin{center}
\resizebox{2.75in}{!}{\includegraphics{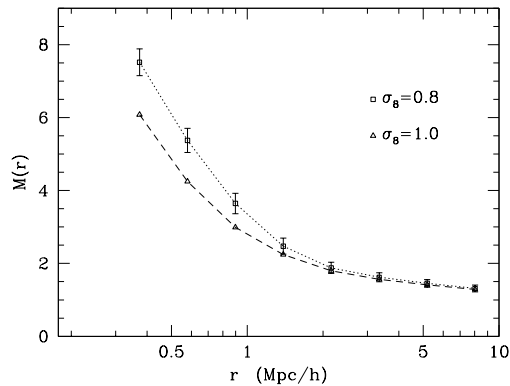}}
\end{center}
\caption{The marked correlation function, $M(r)$, for two HODs which provide
good fits to $\xi(r)$ in the two cosmologies, $\sigma_8=0.8$ (open squares)
and $\sigma_8=1.0$ (open triangles).  The mark is $\rho/(\rho_\star+\rho)$
with $\rho$ determined by kernel estimation using 4 neighbors and
$\rho_\star=25\,\bar{\rho}$ (see text).  The error bars show the diagonal
of the covariance matrix computed for the fiducial model.
When $M(r)$ deviates from $1$, the clustering is sensitive to the (local)
density: we see small scale clustering is enhanced in regions of high
density (as expected) and by different amounts in the two models.}
\label{fig:mark}
\end{figure}

We now measure the local-density marked correlation function for our two
examples HODs.  To begin we imagine that we can use spectroscopy or multi-band
photometry to select a sample of galaxies in a slice $\pm 50\,h^{-1}$Mpc.
At our fiducial $z\simeq 0.1$, or $\chi_\star\simeq 300\,h^{-1}$Mpc, this
corresponds to $\Delta z/(1+z)\sim 15\%$.
In this 2D slice we estimate the density using spline kernel interpolation
with 4 nearest (in projection) neighbors.
Not surprisingly, we find that this density is much higher for objects which
live in massive halos than for those which live in smaller halos.
As the width of the slice is increased the contrast in density between high
and low mass halos is reduced, but the trend remains the same. 
Note that our goal is not to optimize the density estimator, but to 
demonstrate that a useful estimator may be computed even for 
samples with limited redshift information. Of course, the exact choice of 
estimator will depend on the data set being considered.

To pick a reasonable value of $\rho_\star$ we note that halos of
$10^{15}\,h^{-1}M_\odot$ host $\mathcal{O}(10)$ galaxies in our models and
cover $\sim 5\,(h^{-1}{\rm Mpc})^2$ in projection.
Projected over $\pm 50\,h^{-1}$Mpc the background density is
$\sim 0.1\,(h^{-1}{\rm Mpc})^{-2}$, so massive halos are $\sim 20$ more
dense than the mean ($\bar{\rho}$).
We pick $\rho_\star=25\,\bar{\rho}$ as a convenient round number, though our
conclusions are not sensitive to the exact choice.

Figure \ref{fig:mark} shows that this marked correlation function on sub-Mpc
scales is different for our two samples reflecting the differences in the HOD
(Fig.~\ref{fig:hod}).
How discriminatory is this measure?
For our fiducial volume, $\Delta\chi^2\simeq 33$ for the two marked correlation
functions, compared with $\Delta\chi^2 < 1$ for the unweighted correlations.
Since almost all of the difference comes from the lowest 4 data points, the
two models can be strongly discriminated ($>99\%$ assuming Gaussian errors).
The distribution of the marks is almost the same in the two samples, and the
difference in $M(r)$ remains even if we rescale the marks in one model to
match the distribution in the other, showing that the difference is robust.
We also note that the relevant measure of error for $M(r)$ comes from the
Monte-Carlo estimation of the covariance matrix.  Simply scrambling the marks
allows us to test for a density dependence of the correlation function
\protect\citep{SheConSki05}, which is detected in all of our catalogs at
extremely high significance, but does not tell us how to compare different
$M(r)$ to each other. 

Our initial choice of slice width, $\pm 50\,h^{-1}$Mpc, was possibly
optimistic for surveys at higher redshift.  As we increase the width of the
slice the density contrast decreases and the significance by which we can
differentiate the models is also decreased.
For a slice $\pm 125\,h^{-1}$Mpc in width, i.e.~the full depth of our fiducial
$(250\,h^{-1}{\rm Mpc})^3$ survey, using the same mark as above, the two models
in Figure \ref{fig:mark} differ by $\Delta\chi^2=19$.
It is easily conceivable that a different choice of $\rho_\star$ or a higher
power of $\rho$ in the mark could increase the discriminatory power of $M(r)$,
but this is already reasonably significant given that only the 4 points with
$r<1\,h^{-1}$Mpc contribute.  If it proves impossible to select slices as thin
as $\pm 125\,h^{-1}$Mpc for some particular sample, it is always possible to
jointly analyze samples using cross-correlation marks where one of the samples
can be well isolated in distance (e.g.~red galaxies with strong breaks).  Such
statistics would need to be analyzed on a case-by-case basis.

\section{Conclusions} \label{sec:conclusions}

Although the galaxy two-point correlation function has proved to be extremely
useful in modeling the relationship between galaxies and dark matter, it 
does not exhaust the information in the data.
One degeneracy that cannot be broken by the correlation function alone is
between the HOD and cosmology - within the context of the correlation function,
one is free to re-apportion galaxies to compensate for differences in the halo
mass function. The number of such degeneracies will only increase as we attempt
more detailed mappings of galaxies to dark matter in the future.
This note provides proof of principle that such degeneracies can be lifted by
marked correlation functions.

An important advantages of marked correlations is that they do not involve
multiple data sets.  Indeed, for the example presented here, the mark required
{\it no\/} additional information beyond the spatial distribution of galaxies
(and survey mask), which one needed to compute the correlation function in
the first place.
Using the same data set considerably simplifies any modeling step.
Furthermore the marked correlation function can be computed with the same
code and at the same time as the standard correlation function, for which
optimized algorithms exist.
This is a non-negligible advantage when one considers the need to repeatedly
compute it for mock samples while modeling, estimating errors, etc.


Rapid advances in computational power and algorithm development have made
it reasonably straightforward to simulate the distribution of dark matter
halos in large volumes for almost any cosmological model.
Combined with a halo occupation approach this makes ``forward modeling'' of
almost any galaxy statistic possible.
This implies that statistics which use all of the galaxy information, in the
manner of standard large-scale structure $n$-point statistics, can be just as
useful as those which try to identify special subsets of galaxies
(`cluster' vs.~`field').
As our ability to model a wider array of observations (e.g.~weak lensing,
Sunyaev-Zel'dovich decrement or X-ray flux) matures, similar methods can be
applied to these observations, bypassing the need to relate particular
features in a map with individual 3D structures.

\section*{Acknowledgments}

We thank Ravi Sheth and Ramin Skibba for helpful comments on an earlier draft.
MW is supported by NASA and the DoE.
NP is supported by NASA Hubble Fellowship 
NASA HST-HF-01200.01 and an LBNL Chamberlain Fellowship.
The simulations used in this paper were analyzed at the National Energy
Research Scientific Computing Center.
This work was supported by the Director, Office of Science, of the U.S. 
Department of Energy under Contract No. DE-AC02-05CH11231.

\label{lastpage}
\end{document}